\documentclass[aps,pra,superscriptaddress,twocolumn,floatfix]{revtex4}
\usepackage{graphicx}
\usepackage[svgnames]{xcolor}
\usepackage{amsmath,amsfonts,amssymb,mathtools} 
\usepackage[colorlinks=true,allcolors=blue]{hyperref} 
\usepackage[section]{placeins}

\newcommand{\tx}[1]{\text{#1}}

\newcommand{\dagg}[1]{#1^\dagger}
\newcommand{\tr}{\mathrm{tr}}
\DeclarePairedDelimiter{\abs}{\lvert}{\rvert}

\DeclarePairedDelimiter{\ket}{\lvert}{\rangle}
\DeclarePairedDelimiter{\bra}{\langle}{\rvert}
\DeclarePairedDelimiter{\braket}{\langle}{\rangle}
\newcommand{\nbar}{\bar{n}}
\newcommand{\adag}{a^\dagger}
\newcommand{\sigx}{\sigma_x}

\newcommand{\sigz}{\sigma_z}
\newcommand{\sigp}{\sigma_+}
\newcommand{\sigm}{\sigma_-}

\begin{document}

\title{Thermal transport through a single trapped ion under strong laser illumination}

\author{T. Tassis}
\affiliation{Centro de Ci\^encias Naturais e Humanas, Universidade Federal do ABC (UFABC), Santo Andr\'e, SP, 09210--580, Brazil}

\author{F. Brito}
\affiliation{Instituto de F\'isica de S\~ao Carlos, Universidade de S\~ao Paulo, S\~ao Carlos (SP), Brazil}
\affiliation{Quantum Research Centre, Technology Innovation Institute, P.O. Box 9639, Abu Dhabi, United Arab Emirates}

\author{F. L. Semi\~ao}
\thanks{Corresponding author: fernando.semiao@ufabc.edu.br}  
\affiliation{Centro de Ci\^encias Naturais e Humanas,  Universidade Federal do ABC (UFABC), Santo Andr\'e, SP, 09210--580, Brazil}

\begin{abstract}
In this work, we study quantum heat transport in a single trapped ion, driven by laser excitation and coupled to thermal reservoirs operating at different temperatures.
Our focus lies in understanding how different laser coupling scenarios impact the system dynamics.
As the laser intensity reaches a regime where the electronic and motional degrees of freedom of the ion couple strongly, traditional approaches using phenomenological models for thermal reservoirs become inadequate.
Therefore, the adoption of the dressed master equation (DME) formalism becomes crucial, enabling a deeper understanding of how distinct laser intensities influence heat transport.
Analyzing the heat current within the parameter space defined by detuning and coupling strength, we observe intriguing circular patterns which are influenced by the vibrational frequency of the ion and laser parameters, and reveal nuanced relationships between heat transport and coherence, as well as phenomena such as negative differential heat conductivity and heat rectification, offering insights into the thermal properties of this essential quantum technology setup.
\end{abstract}

\maketitle

\section{Introduction}

The desire to build universal quantum computers \cite{nielsen2010} has led to a remarkable development of methods for controlling quantum systems and, additionally, the creation of increasingly smaller and more complex quantum devices.
In order to have the level of control necessary for this application, it is important, among other things, to understand how these quantum systems interact with their environment \cite{breuer2002}.
One of the efforts in this direction is that of understanding nonequilibrium processes, e.g., transport of energy, and how the response changes according to system or reservoir properties.

On the other hand, few quantum technologies are as advanced as the setup involving trapped ions interacting with lasers and cavity fields \cite{leibfried2003,haffner2008,bruzewicz2019,css}.
Ion traps enable precise control over state initialization, dynamics, and system measurements.
The possibility of arranging the ions in different spatial configurations, including 2- and 3d crystals, along with laser driving, opens the door to simulating a vast number of condensed matter systems \cite{blatt2012,lamata2014,joshi2020,monroe2021}.
Additionally, the capability to engineer reservoirs renders the trapped ion systems ideal setups for investigating quantum thermodynamic cycles \cite{teixeira2022}.
Understanding transport in such controllable circumstances can advance not only the theory of out-of-equilibrium systems and many-body physics, but also provide insights for the development of new technologies  \cite{joshi1993,lepri2003,manzano2012,asadian2013,bermudez2013,freitas2015,nicacio2015,joulain2016,manzano2016,nicacio2016,guo2019,maier2019,meher2020,chen2022,yang2020,landi2022,yan2023}.

In this work, we are interested in exploring the transport response in varied coupling scenarios, linked to the variation of the laser intensity employed for manipulating the trapped ion.
This variation of the coupling strength makes it imperative to use the dressed master equation (DME) formalism to achieve accurate, physical results \cite{scala2007,beaudoin2011,santos2014,levy2014,chiara2018}.
In particular, we find an interesting relation approximately satisfied by the laser-ion coupling constant $\Omega$ and detuning $\delta$, as well as the trap frequency $\nu$.
This relation reads $\Omega^2 + \delta^2 = (m \nu)^2$, and its fulfillment approximately gives the optimized current.
Surprisingly, this relation is also related to local maxima or minima of the leftover coherence in the steady state.
Additionally, we show the controlled emergence of negative differential heat conductivity \cite{yomo2005,elste2006,li2006,benenti2009,he2010} in this system, a phenomenon characterized by a nonmonotonic behavior of the current with respect to temperature difference between the two reservoirs.
Finally, we studied the asymmetric character of the current \cite{motz2018,simon2019,kalantar2021} with respect to the swap of reservoirs.
In particular, we studied how the rectification factor responds to controlled parameter changes.

The paper is organized as follows.
In Sec.~\ref{sec:model} we present the model we are going to consider, giving a
brief review of the theoretical description of trapped ions, the treatment of
the open quantum system via the DME, and how to obtain the heat current.
We proceed to present some numerical results in Sec.~\ref{sec:results} in a wide range of physical parameters.
Finally, in Sec.~\ref{sec:conclusion} we summarize the results and make some final remarks.

\section{Model}\label{sec:model}

We are interested in studying the properties of a single trapped ion coupled to thermal reservoirs at different temperatures.
A trapped ion can be effectively characterized by its internal electronic state and the position of its center of mass.
Through the application of selection rules and appropriate detunings, the electronic subspace can be simplified into a two-level system \cite{leibfried2003}.
Similarly, by adjusting the electromagnetic trapping fields, the motion of the ion's center of mass can be accurately portrayed as a harmonic oscillation around an equilibrium point along the trap axis \cite{leibfried2003}.
The application of the laser induces coupling between the electronic and motional parts, and the Hamiltonian describing their interaction can be cast in the form \cite{moya-cessa2012}
\begin{equation}\label{sH}
  H_S = \nu \adag a + \frac{\delta}{2} \sigz
  + \frac{\Omega}{2} \bigl[ \sigp e^{i \eta (a + \adag)}
  + \sigm e^{-i \eta (a + \adag)} \bigr],
\end{equation}
where $\nu$ is the ion's vibrational angular frequency, $\delta := \omega_0 - \omega_L$ is the detuning between the electronic transition frequency $\omega_0$ and the laser frequency $\omega_L$, $\Omega$ is the coupling constant, and $\eta$ is the Lamb-Dicke (LD) parameter.
The operator $a$ ($\adag$) annihilates (creates) a vibration quantum, $\sigm$ ($\sigp$) is the spin lowering (raising) operator, and $\sigz$ is the $z$ Pauli operator, which also acts on the spin subsystem. In this equation, and throughout the following, we adopt natural units where, in particular, the reduced Planck constant $\hbar$ (i.e., $h$ over $2\pi$) and the Boltzmann constant $k_B$ are set to one.

\begin{figure}[!tbh]
    \centering
    \includegraphics[width=0.48\textwidth]{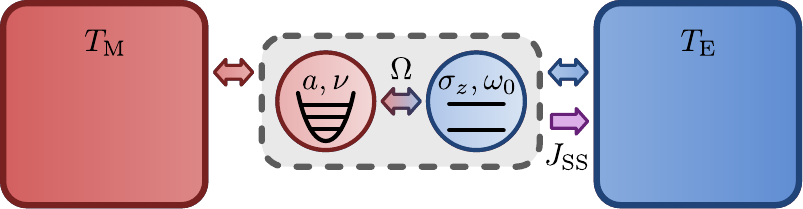}
    \caption{(Color online) Schematic illustration of the system in consideration.
A trapped ion (gray, dashed-border rectangle) can be described by its motional (M) and electronic (E) subsystems, represented by the left red circle and the right blue circle, respectively.
These can be approximated, respectively, by a quantum harmonic oscillator of frequency $\nu$ and a two-state system of energy split $\omega_0$.
The motional and electronic subsystems can be coupled via a laser, with the coupling strength denoted by $\Omega$.
By connecting thermal reservoirs at different temperatures $T_M$ and $T_E$ (red and blue, solid-border rectangles) to the motional and electronic parts, respectively, it is expected that the system will reach a stationary state with constant flux of energy given by the heat current $J_{SS}$ (purple arrow).}
    \label{fig:system}
\end{figure}

In this work, we explore the nonequilibrium consequences of introducing a temperature gradient across the ion under different coupling conditions.
Typically, this temperature gradient is achieved through reservoir engineering.
At a theoretical level, we capture the physics by subjecting both the electronic and motional degrees of freedom to independent Markovian baths at a specified temperature $T_E$ and $T_M$, respectively.
Consequently, upon laser illumination, the two subsystems are expected to exchange energy through their laser-induced coupling, eventually settling into an asymptotic state.
The energy current induced by the two thermal baths in this state is of particular interest to us.
As we will see, the magnitude of the current exiting one reservoir equals the one entering the other in the asymptotic state.
This scenario is depicted in Fig.~\ref{fig:system}. 

From an implementation perspective, the thermal bath for the electronic degrees of freedom can be engineered by using dedicated lasers that promote the coupling of the target ion to other auxiliary ions and their common motion in the trap. As detailed in \cite{Duan}, the spectral density of the resulting reservoir depends on various factors, including the ion number, the target ion location, the laser detuning relative to the motional sidebands, and the number of frequency components in the laser. Concerning the vibrational motion, laser cooling and heating can be used to implement effective thermal reservoirs. The interested reader can refer to \cite{teixeira2022} and the references therein. In \cite{Bermudez}, this approach is considered to promote energy transport in phonon-mediated spin–spin interactions in crystals of trapped atomic ions. The physical picture here is that laser cooling couples the vibrational mode to an infinite number of photonic modes in the electromagnetic bath, thereby creating an effective thermal bath for the vibrational degrees of freedom.

The reservoirs are included in the dynamical description of the system by means of the bath Hamiltonian \cite{chen2022}
\begin{equation}\label{eq:hamilt_baths}
    H_B = \sum_{\mu \in \{E, M\}} H_{B \mu}^0 + V_\mu,
\end{equation}
where
\begin{equation}
H_{B \mu}^0 = \sum_k \omega_{\mu k} \dagg{b}_{\mu k} b_{\mu k}
\end{equation}
is the free Hamiltonian of the $\mu$-th reservoir, and
\begin{equation} \label{eq:bath-V}
    V_\mu = A_\mu \otimes \sum_k g_{\mu k} (b_{\mu k} + \dagg{b}_{\mu k})
\end{equation}
is the respective interaction term with the subsystem $\mu \in \{E, M\}$. We consider that the system-reservoir interaction is weak, and that reservoir correlations decay much faster than any significant time scale of the system.
In these conditions, we can use the usual Born-Markov approximations.
Regarding the specific form of the system operators in Eq.~(\ref{eq:bath-V}), we will be considering $A_M = a + \adag$ and $A_E = \sigx$, which is a common choice to describe energy exchange with thermal baths. In this way, the dynamics of the system is given by the reduced density operator, $\rho := \tr_B\{\rho_{SB}\}$, which obeys the dressed master equation (DME) \cite{breuer2002,scala2007,beaudoin2011}
\begin{equation}\label{eq:dme}
    \frac{d \rho}{dt} = - i [H_S, \rho] + \sum_{\mu jk} \Gamma_{\mu jk} D_{jk}[\rho],
\end{equation}
where
\begin{equation} \label{eq:Gamma}
    \Gamma_{\mu jk} = \begin{cases}
    \gamma_{\mu jk} \, \nbar_\mu(\omega_{jk}) \, \abs{\braket{k | A_\mu | l}}^2, \tx{ if } \omega_{jk} > 0 \\
    \gamma_{\mu jk} \, [1 + \nbar_\mu(- \omega_{jk})] \, \abs{\braket{k | A_\mu | l}}^2, \tx{ if } \omega_{jk} < 0
    \end{cases},
\end{equation}
and
\begin{equation}
    D_{jk}[\rho] = P_k \ket{j} \bra{j} - \frac12 \ket{k} \bra{k} \rho - \frac12 \rho \ket{k} \bra{k}
\end{equation}
Here, $\ket{k}$ is the eigenvector of $H_S$ such that $H_S \ket{k} = E_k \ket{k}$, $\omega_{jk} = E_j - E_k$, $P_k = \braket{k | \rho | k}$, $\nbar_\mu(\omega_{jk}) = (e^{\omega_{jk}/T_\mu} - 1)^{-1}$, and $\gamma_{\mu jk}$ are constants, which depend on particularities of the chosen baths. Unless explicitly stated, we consider equal relaxation rates $\gamma_{\mu j k} = \gamma$. Only when analyzing the phenomenon of current rectification do we consider a case with different such rates.

Before proceeding, it is important to justify our use of the DME instead of the more common phenomenological, local master equation approach. To derive, for example, a local dissipator of the form \cite{scully}
\begin{equation}
    D[\rho] = \gamma \biggl( a \rho \adag - \frac12 \{\adag a, \rho\} \biggr)
\end{equation}
for the vibrational motion, one must neglect its coupling to any other quantum system, except its own bath. In our case, this would correspond to neglecting the electronic degree of freedom. When the coupling constant $\Omega$ becomes much stronger than the decay rate $\gamma$, the local master equation formalism begins to break down, as the system is no longer separable. For instance, the reservoir coupled to the motional part becomes indirectly coupled to the electronic part via the laser-induced interaction between them.

When a system is placed in contact with reservoirs at different temperatures, it is expected not to reach an equilibrium state but rather a time-independent, asymptotic stationary state. In this state, the total rate of energy change is zero, yet it can be split into contributions with opposite signs, interpreted as a constant energy flux from one reservoir to the other  \cite{asadian2013,nicacio2016}. The starting point to find the heat current due to each of the reservoirs is to notice that the total current reads

\begin{align}
    J &\coloneqq - \frac{d}{dt} \braket{H_S} = - \tr\biggl[\frac{d \rho}{dt} H_S \biggr] \\
    &= - \sum_{\mu jk} \omega_{jk} \, \Gamma_{\mu jk} \, P_k, \label{eq:curr-both}
\end{align}
where the minus sign indicates that the current is considered positive when energy leaves the system. We can now split the summation in $\mu \in \{E, M\}$ as $J = J_E + J_M$ to obtain the desired current for each reservoir. In particular, for the steady state denoted as $\rho_{SS}$, we have $d \rho_{SS} / dt = 0$ and, consequently, $J = 0$.
In this scenario, we can write
\begin{equation} \label{eq:curr}
    J_{SS} \coloneqq J_E = \sum_{jk} \omega_{jk} \Gamma_{E jk} P_k = - J_M,
\end{equation}
where $J_E$ is the current between the electronic subsystem and its bath and $J_M$ is the analogous current for the center-of-mass subsystem and its bath. As noted above, both currents are equal in magnitude in the steady state.

\section{Results} \label{sec:results}

This section presents numerical simulations of the system in Fig.~\ref{fig:system} responding to a temperature gradient across various parameter regimes.
All of the numerical results were obtained with the use of the QuTiP framework \cite{johansson2012} for Python.
We started by calculating $\rho_{SS}$ from the DME in Eq.~(\ref{eq:dme}) and proceeded to calculate the current using Eq.~(\ref{eq:curr}).
For all the results presented here, the Lamb-Dicke (LD) parameter was set to $\eta = 0.05$ and the dimension of the space of states of the motional bosonic mode (frequency $\nu$) was truncated at $N=30$ (truncated Fock basis).
We experimented with other values of the LD parameter inside the LD regime, $\eta \ll 1$, and did not observe significant differences in the qualitative aspects of the results presented here.
We also experimented with positive and negative values of the detuning $\delta$.
However, the current magnitudes seem to only depend on $\abs{\delta}$, with the current $J_{SS}$ varying less than $0.1\%$ between positive and negative detunings.
Thus, without any generality loss, we will present only the results for positive $\delta$.

\begin{figure}[!tbh]
    \centering
    \includegraphics[width=0.36\textwidth]{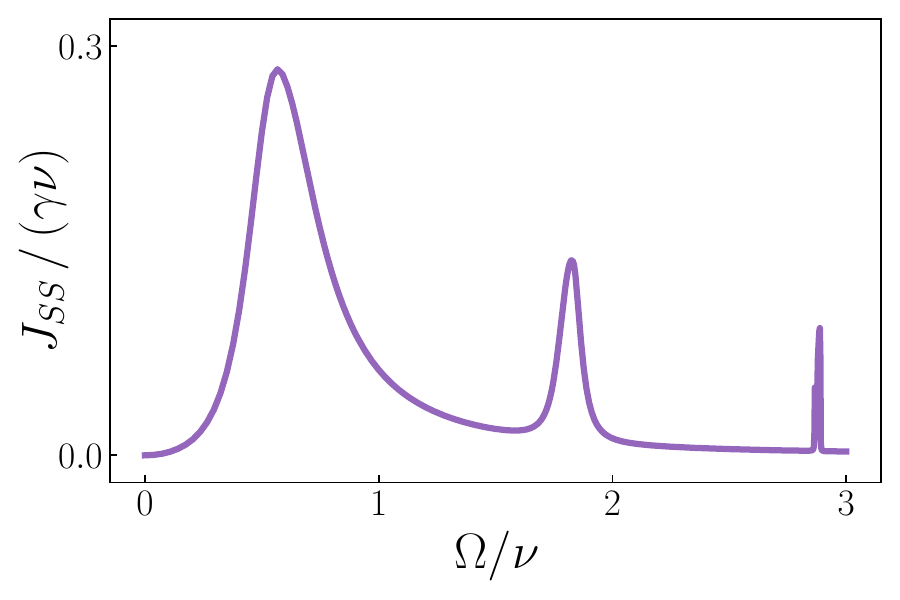}
    \includegraphics[width=0.36\textwidth]{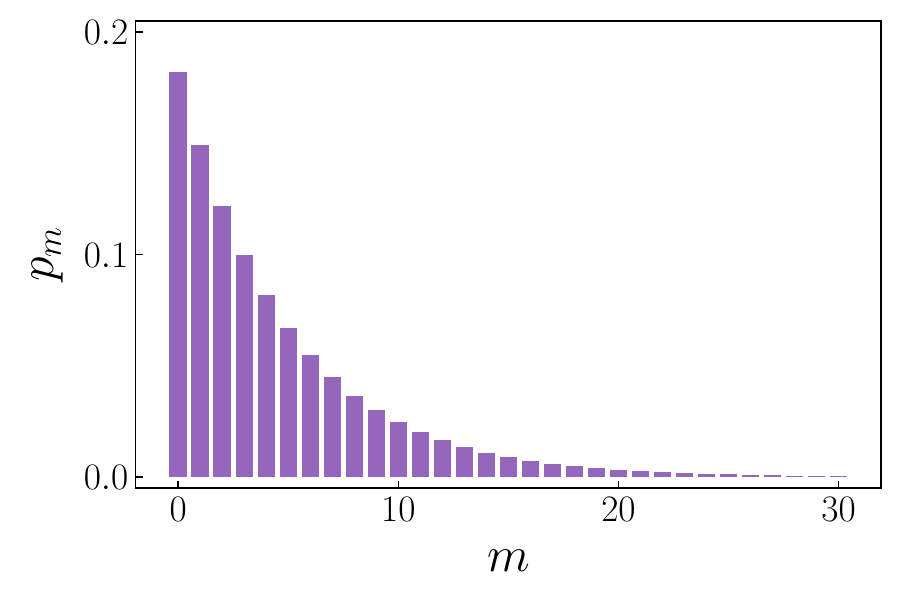}
    \caption{
       Top panel: Steady-state heat current $J_{SS}$ as a function of the coupling strength $\Omega$ for a fixed detuning $\delta = 0.8\nu$. The reservoir temperatures were set to $T_E = 0.5 \nu$ and $T_M = 5 \nu$. Bottom panel: Steady-state population of the 30 states in the truncated Fock basis $|m\rangle$ (with $0 \leq m \leq N = 30$) for the bosonic motional mode, corresponding to the vibrational degree of freedom of the trapped ions (with frequency $\nu$). This plot was obtained with $\Omega = 1.5 \nu$, but similar results are obtained for all values of $\Omega$ considered in the top panel plot.
    }
    \label{fig:curr}
\end{figure}

Figure~\ref{fig:curr} shows simulation results for reservoirs at temperatures $T_E = 0.5\nu$ and $T_M = 5\nu$, with a fixed detuning of $\delta = 0.8\nu$. The top panel displays the normalized current, $J_{SS}/(\gamma \nu)$, as a function of the coupling constant $\Omega$. Notably, the behavior of $J_{SS}$ is highly nonmonotonic with respect to $\Omega$, exhibiting a series of current peaks interspersed with regions of current suppression. The bottom panel shows the steady-state population $p_n$ in the truncated Fock basis used to represent the vibrational state in the simulations for a specific value of the Rabi frequency. It shows that the chosen number of elements in the basis, $N = 30$, is sufficient to yield accurate numerical results.

\begin{figure}[!tbh]
    \centering
    \includegraphics[width=0.48\textwidth]{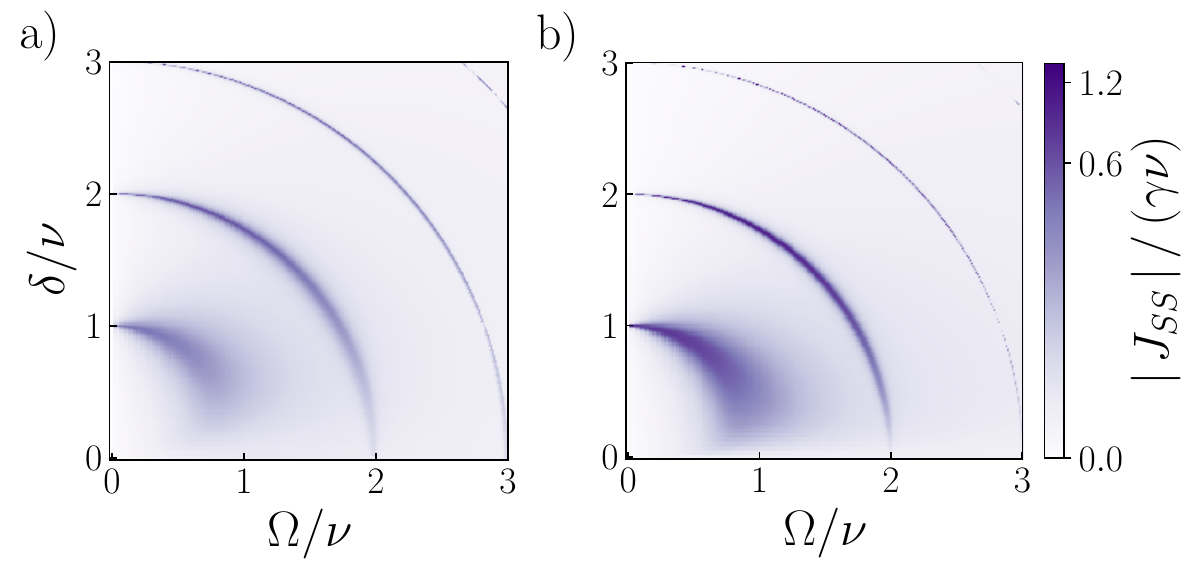}
    \caption{
      (Color online) Magnitude of the steady-state heat current $J_{SS}$
      as a function of the detuning $\delta$ and the coupling strength $\Omega$.
      (a) $T_E = 0.5 \nu$ and $T_M = 5 \nu$.
      (b) $T_E = 5 \nu$ and $T_M = 0.5 \nu$.
    }
    \label{fig:curr2d}
\end{figure}

In Fig.~\ref{fig:curr2d},
we extrapolate the previous simulation for several values of $\delta$
and present the normalized current as a function of both $\delta$ and $\Omega$.
For Fig.~\ref{fig:curr2d}a,
the electronic and motional reservoirs were set, respectively,
at temperatures $T_E = 0.5 \nu$ and $T_M = 5.0 \nu$,
while for Fig.~\ref{fig:curr2d}b, we have the inverse scenario.
In both cases, the maximum values of the current, the darker areas,
form circular patterns in the $(\Omega, \delta)$ space.
Additionally, the currents are asymmetric, with stronger maxima when the electronic part is coupled to the hot reservoir.
This asymmetry will be further explored shortly.

\begin{figure}[!tbh]
    \centering
    \includegraphics[width=0.48\textwidth]{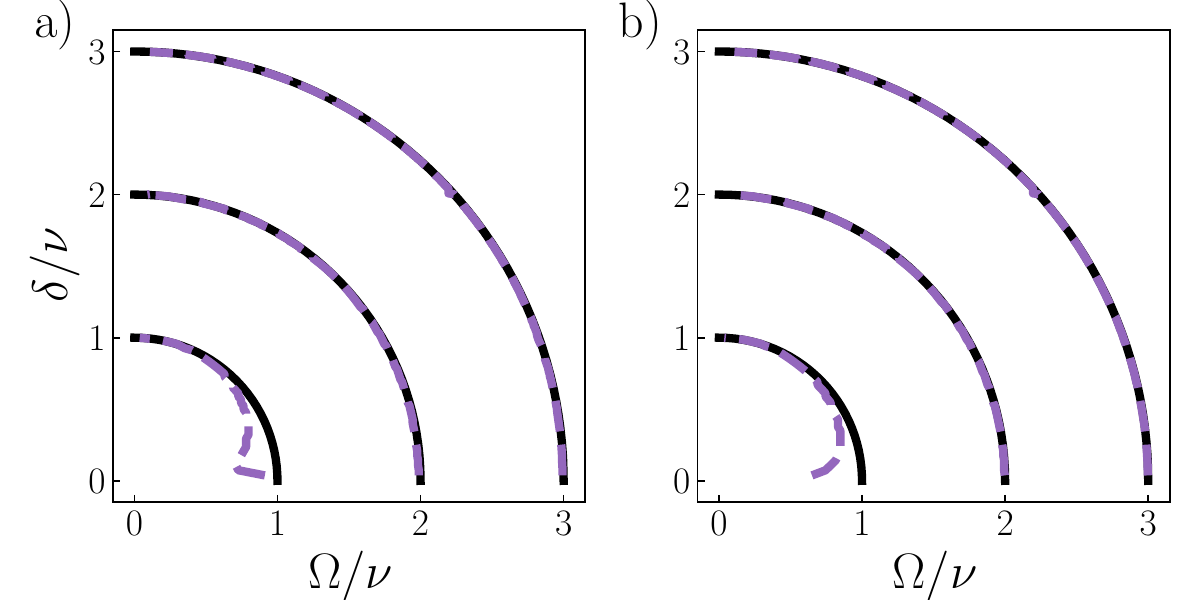}
    \caption{
      (Color online) Location of the local maxima (at constant $\delta$)
      of $\abs{J_{SS}}$ (dashed purple lines) in the $(\Omega, \delta)$ space.
      The circular sectors,
      $\delta^2 + \Omega^2 = (m \nu)^2$, $m=1,2,3$,
      are depicted as solid black lines.
      (a) $T_E = 0.5 \nu$ and $T_M = 5 \nu$.
      (b) $T_E = 5 \nu$ and $T_M = 0.5 \nu$.
    }
    \label{fig:circles}
\end{figure}

We now take a closer look at the appearance of the circular patterns. For each fixed value of $\delta$, we find the critical $\Omega$ that satisfies $dJ_{SS}/d\Omega = 0$. The results are shown in Fig.~~\ref{fig:circles}, where we can see that these local maxima tend to occur close to the circular sectors
\begin{equation}\label{circular}
\delta^2 + \Omega^2 = (m \nu)^2,
\end{equation} 
where $m = 1, 2, \dots$ represents a natural number. This analysis confirms that, with the exception of a few isolated regions, particularly for $m = 1$, such maxima of the current align closely with these circular sectors, as anticipated in Fig.~\ref{fig:curr2d}, confirming a significant relationship between the parameters.

Interestingly, it was shown recently \cite{tassis2023} that along the first circular sector, $\delta^2 + \Omega^2 = \nu^2$, the trapped ion Hamiltonian can be approximated, in a unitarily rotated frame, to that of the Jaynes-Cummings model. While this result might appear to simplify the problem entirely, suggesting that the dynamics being fully described by the Jaynes-Cummings model would naturally lead to simple or intuitive analytical expressions for the heat current, this is not the case. Although the Jaynes-Cummings model is exactly solvable, its emergence in the present setup under the resonance $\delta^2 + \Omega^2 = \nu^2$ occurs within a nontrivial transformed frame \cite{tassis2023}. This picture relies on carefully chosen local and bipartite unitary transformations, leading to complex calculations even for simple observables. In particular, the expression for the heat current in Eq.~(\ref{eq:curr-both}) becomes a highly intricate form, comprising infinite sums of non-intuitive and cumbersome terms.

Regarding the other circular sectors [Eq.~(\ref{circular}) with $m \geq 2$], the underlying physical mechanism for their emergence is expected to be similar to the case of $m = 1$, reported in \cite{tassis2023}. Specifically, these circular conditions $\delta^2 + \Omega^2 = (m \nu)^2$ are likely associated with the emergence of effective Hamiltonians that involve $m$ phonon transitions [$a^m$ and $(a^\dag)^m$], leading to  generalizations of the Jaynes-Cummings model. This is consistent with the fact that $\nu$ denotes the vibration frequency (phonon mode). Deriving effective Hamiltonians for these cases requires identifying the appropriate unitarily transformed frames where the effective dynamics arise upon resonance, as discussed in  \cite{tassis2023} for $m=1$. To the best of our knowledge, no such frames or effective Hamiltonians have yet been reported for $m \geq 2$, which makes this an interesting research problem by itself.

Another important feature of this plot is that the widths of the regions where the current is appreciable, specifically around $\delta^2 + \Omega^2 = (m \nu)^2$, become narrower as $m$ increases. This trend can also be seen in the top panel of Fig.~\ref{fig:curr}. The key to understanding this behavior again lies in the $m\text{-phonon}$ transitions, represented by $a^m$ and $(a^\dag)^m$. These transitions arise from an expansion in $\eta$ of the exponential factor $\exp[\pm i\eta (a + a^\dag)]$, which appears in the interaction part of the Hamiltonian in Eq.~(\ref{sH}). The contribution from the $m\text{-phonon}$ terms is scaled by a factor of $\eta^m$. Since $\eta$ is typically much smaller than one, the resonance condition $\delta^2 + \Omega^2 = (m \nu)^2$ corresponds to relatively weak $m\text{-phonon}$ transitions, which lead to sharp lines. This sharpness occurs because, as soon as we deviate from the exact values of $\delta$ and $\Omega$ that satisfy the resonance condition, the dynamics are significantly perturbed. This explains the observed narrowness of the lines.

Another interesting result connected to the circles in Eq.~(\ref{circular}) relates to the coherence that still persists in the asymptotic state as the result of the interplay between coherent (Hamiltonian) and incoherent (thermal) dynamics. To quantify this quantum resource, we choose the relative entropy of coherence evaluated in the free basis $\mathcal{F}=\{ \ket{e,n}, \ket{g,n} \}$. It reads \cite{baumgratz2014}
\begin{equation} 
  C(\rho_{SS}) = S(\rho_{SS}^\mathrm{diag}) - S(\rho_{SS}),
\end{equation}
where $S$ is the von Neumann entropy, and $\rho_{SS}^\mathrm{diag}$ has the same diagonal elements of $\rho_{SS}$ in the $\mathcal{F}$ basis and zeros in all other positions.

\begin{figure}[!tbh]
    \centering
    \includegraphics[width=0.48\textwidth]{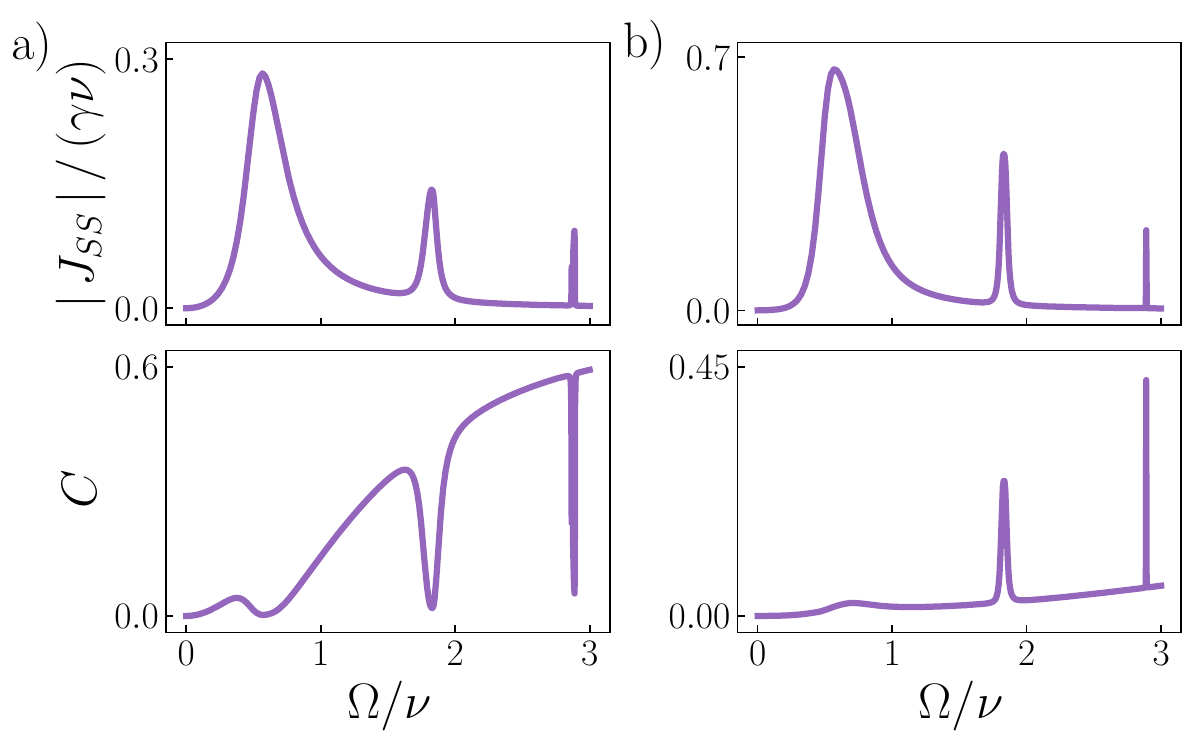}
    \caption{
      (Color online) Magnitude of the steady-state current $J_{SS}$
      and relative entropy of coherence $C$ of the steady state,
      calculated in the free Hamiltonian basis $\mathcal{F}$,
      as a function of $\Omega$ and for fixed $\delta = 0.8 \nu$.
      (a) $T_E = 0.5 \nu$ and $T_M = 5 \nu$.
      (b) $T_E = 5 \nu$ and $T_M = 0.5 \nu$.
    }
    \label{fig:coh}
\end{figure}

  In Fig.~\ref{fig:coh}, we present a comparison of the relative entropy of
  coherence $C(\rho_{SS})$ and the steady-state current $J_{SS}$
  as a function of $\Omega$ and for fixed $\delta = 0.8 \nu$.
  In Fig.~\ref{fig:coh}a we have temperatures $T_E = 0.5\nu$ and $T_M = 5 \nu$,
  while in Fig.~\ref{fig:coh}b, we have $T_E = 5\nu$ and $T_M = 0.5 \nu$.
  We observe an intriguing correlation between current and coherence. In the first case, there appears to be a negative correlation, with coherence decreasing as current increases. In contrast, the second case exhibits a positive correlation, where peaks in the current align with peaks in coherence. Once again, we observe a pronounced asymmetry regarding the direction of the temperature gradient.

\begin{figure}[!tbh]
    \centering
    \includegraphics[width=0.48\textwidth]{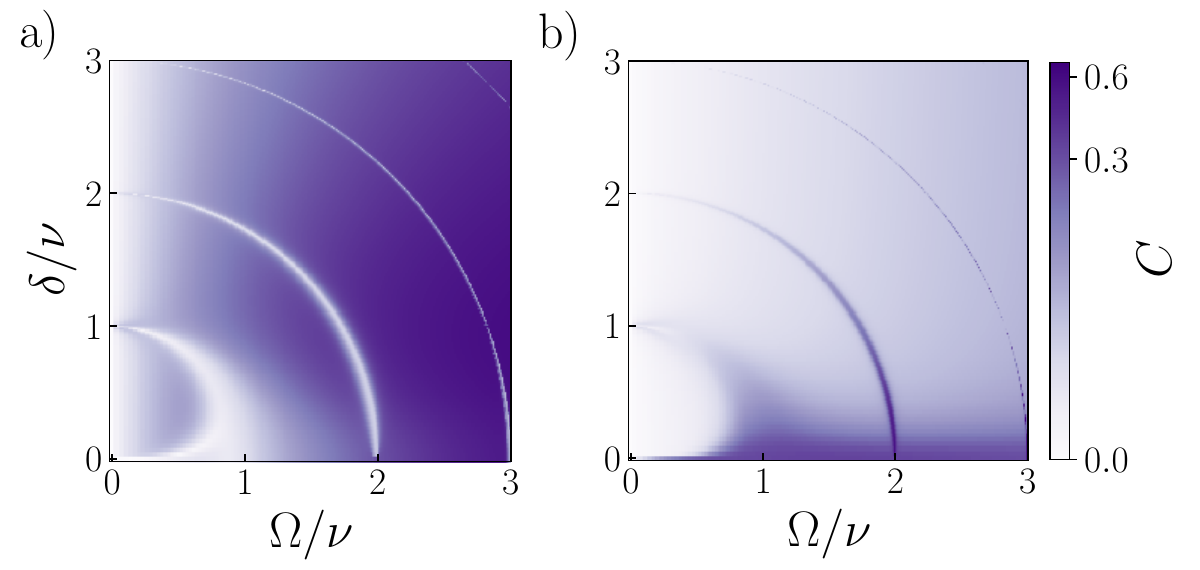}
    \caption{
      (Color online) Relative entropy of coherence $C$
      of the steady state,
      calculated in the free Hamiltonian basis $\mathcal{F}$,
      as a function of $\delta$ and $\Omega$.
      (a) $T_E = 0.5 \nu$ and $T_M = 5 \nu$.
      (b) $T_E = 5 \nu$ and $T_M = 0.5 \nu$.
    }
    \label{fig:coh2d}
\end{figure}

To further explore the relationship between current and coherence in the asymptotic state, we plot the coherence as a function of $\delta$ and $\Omega$, following the approach used for the current in Fig.~\ref{fig:curr2d}. The results are presented in Fig.~\ref{fig:coh2d}.
Similar to the current, the figures reveal circular patterns in this parameter space. Notably, the correlation between current and coherence observed in Fig.~\ref{fig:coh}
appears to be consistent.
Specifically, when the electronic component is coupled to the cold reservoir
(Fig.~\ref{fig:coh2d}a), peaks in the current correspond to abrupt decreases in the relative entropy of coherence. Conversely, in the opposite scenario (Fig.~\ref{fig:coh2d}b), current peaks are associated with increases in the coherence present in the steady state.

Having established sound connections between coherence and current, we now deepen our understanding of the current by studying how the trapped ion behaves in response to varying temperature gradients.
While it is generally expected that the current magnitude will increase monotonically with temperature bias, this is not always the case. Some systems exhibit what is known as negative differential conductivity (NDC) \cite{yomo2005,elste2006,li2006,benenti2009,he2010}.

\begin{figure}[!tbh]
    \centering
    \includegraphics[width=0.36\textwidth]{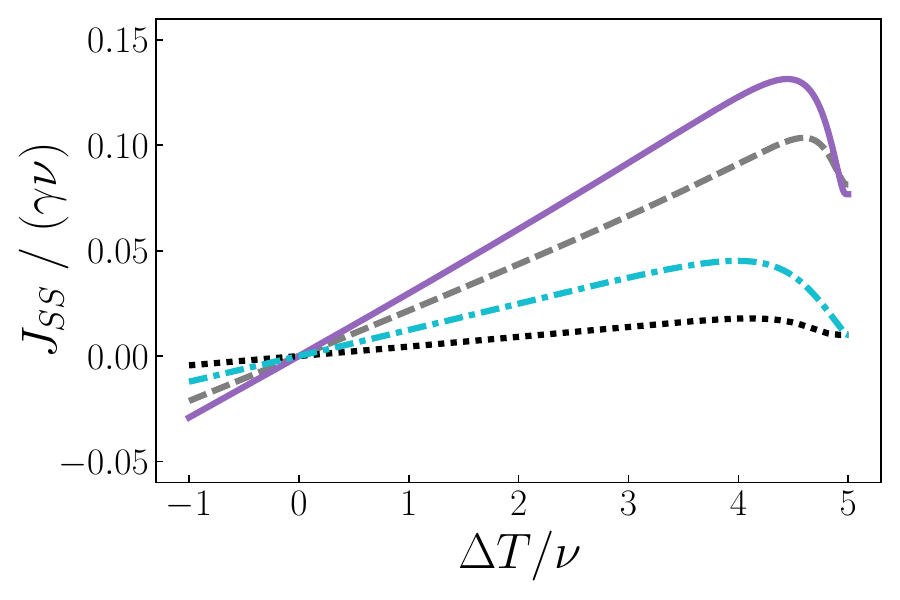}
    \caption{(Color online) Heat current $J_{SS}$ as a function
      of the temperature bias $\Delta T \equiv T_M - T_E$.
      For each line, the coupling strength was set to:
      $\Omega = 0.4 \nu$ (dashed grey);
      $\Omega = 0.8 \nu$ (solid purple);
      $\Omega = 1.5 \nu$ (dotted black);
      and $\Omega = 1.9 \nu$ (dash-dotted cyan)
      The detuning was set to $\delta = 0.8 \nu$.
      The temperature of the motional reservoir was fixed at $T_M = 5\nu$.
    }
    \label{fig:jpeak}
\end{figure}

In the following, we set the temperature of the motional reservoir to $T_M = 5\nu$ and vary the temperature of the reservoir coupled to the electronic part, $T_E$. In Fig.~\ref{fig:jpeak}, we present results for a fixed detuning $\delta = 0.8 \nu$ and different values of the Rabi frequency $\Omega$. In all cases, we observe the onset of NDC, where the current no longer increases as the temperature gradient, $\Delta T \equiv T_M - T_E$, increases. However, for smaller temperature differences, the current exhibits the linear behavior $J \propto \Delta T$, consistent with Fourier’s Law. Similar NDC behavior is observed for different values of $\delta$ and $\Omega$, indicating a consistent response across these parameters. Additionally, when $\Delta T \approx 0$, the current behaves as expected, with heat flowing from the hot to the cold reservoirs, as indicated by the change in the signal of $J_{SS}$.

Negative differential conductivity (NDC) is typically associated with the onset of nonlinearities. This well-established fact is exemplified by the well-known result that coupled harmonic oscillators cannot exhibit NDC \cite{li2006, he2010}. The results shown in Fig.(\ref{fig:jpeak}) can thus be understood as arising from the inherent nonlinearity introduced by the laser-driven two-level system, which leads to the exponential function $\exp[\pm i\eta (a + a^\dag)]$ in the interaction part of the Hamiltonian, as given in Eq.(\ref{sH}). At sufficiently large temperature gradients, higher energy levels become populated, making the contribution of nonlinearity appreciable, leading to the observed NDC.

The next and final aspect of transport we examine in this work is the onset of current rectification \cite{motz2018,simon2019,meher2020,kalantar2021} and its sensitivity to variations in the same parameters as before, which are tuned through the control of the laser — specifically, the detuning $\delta$ and the coupling constant $\Omega$. When the reservoirs are swapped, the current direction reverses, but an asymmetry in current magnitude may emerge, with a stronger flow in one direction.
This can be noted by comparing the magnitudes of $J_{SS}$
in Figs.~\ref{fig:curr2d}a and \ref{fig:curr2d}b.
A quantifier of this asymmetry is found in the so-called rectification factor \cite{meher2020}
\begin{equation}\label{R}
    R \coloneqq \frac{\abs{J_\rightarrow} - \abs{J_\leftarrow}}
    {\max\{\abs{J_\rightarrow}, \abs{J_\leftarrow}\}},
\end{equation}
where $J_\rightarrow$ is the current when the cold reservoir is coupled to the electronic part, and $J_\leftarrow$ is the current in the opposite scenario. From its definition in Eq.~(\ref{R}), we see that the rectification factor is bounded by $\abs{R} \leq 1$. Specifically, $R = 0$ indicates symmetric current flow, while $\abs{R} = 1$ represents complete rectification, where the current flows only in one direction, as in a perfect diode.

\begin{figure}[!tbh]
    \centering
    \includegraphics[width=0.36\textwidth]{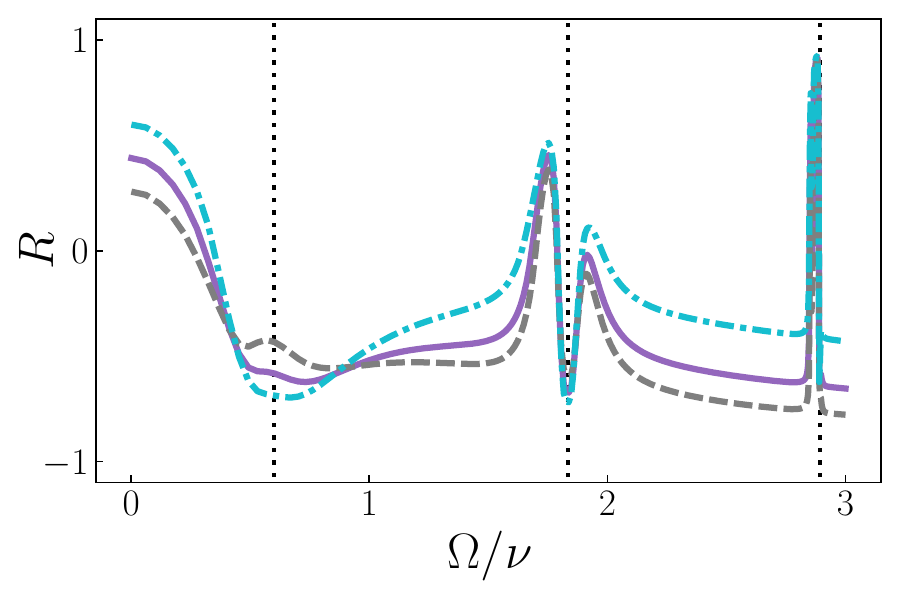}
    \caption{
      (Color online) Rectification factor as a function of the coupling constant $\Omega$ for different decay rates.
      For each line, the decay rate of the reservoir
      coupled to the motional part was set to:
      $\gamma_M = 0.5 \gamma$ (dashed grey);
      $\gamma_M = \gamma$ (solid purple);
      and $\gamma_M = 2 \gamma$ (dash-dotted cyan).
      The decay rate for the electronic reservoir
      was always set to $\gamma_E = \gamma$
      The hot and cold reservoirs were fixed, respectively,
      at temperatures $T_H = 5\nu$ and $T_C = 0.5 \nu$.
    }
    \label{fig:rect}
\end{figure}

Up to this point, we have considered equal relaxation rates for both reservoirs, $\gamma_{E j k} = \gamma_{M j k} = \gamma$, in Eq.(\ref{eq:Gamma}). Next, we examine the effects on the rectification factor of introducing different such rates. In Fig.\ref{fig:rect}, we present the rectification factor $R$ as a function of the coupling strength $\Omega$, for a fixed detuning $\delta = 0.8 \nu$. The three curves shown correspond to a fixed electronic reservoir relaxation rate $\gamma_{E j k} = \gamma$, while varying the relaxation rates for the motional reservoir. Specifically, we considered the cases $\gamma_{M j k} \equiv \gamma_M = 0.5 \gamma$ (dashed grey), $\gamma_M = \gamma$ (solid purple), and $\gamma_M = 2 \gamma$ (dash-dotted cyan). The locations of the circular sectors are marked by dotted vertical lines. It can be seen that the rectification factor is more affected by changes in the relaxation rates in some regions than in others. In particular, in the first circular sector (first dotted vertical line), an increase in the relaxation rate $\gamma_M$ leads to more negative values of $R$. Conversely, in the third circular sector, a decrease in $\gamma_M$ facilitates negative rectification. There are also small regions where the curves cross, which remain robust to small changes in the relaxation rates.

This phenomenology suggests that the response of $R$ to variations in the thermal relaxation rates, combined with the flexibility offered by lasers of different intensities and detunings, opens up avenues for future investigations. These studies could focus on optimizing these parameters to achieve targeted values of the rectification factor. This approach could further broaden the potential applications of this system for thermal devices \cite{kalantar2021} and non-equilibrium thermodynamics \cite{neqt}.

\begin{figure}[!tbh]
    \centering
    \includegraphics[width=0.36\textwidth]{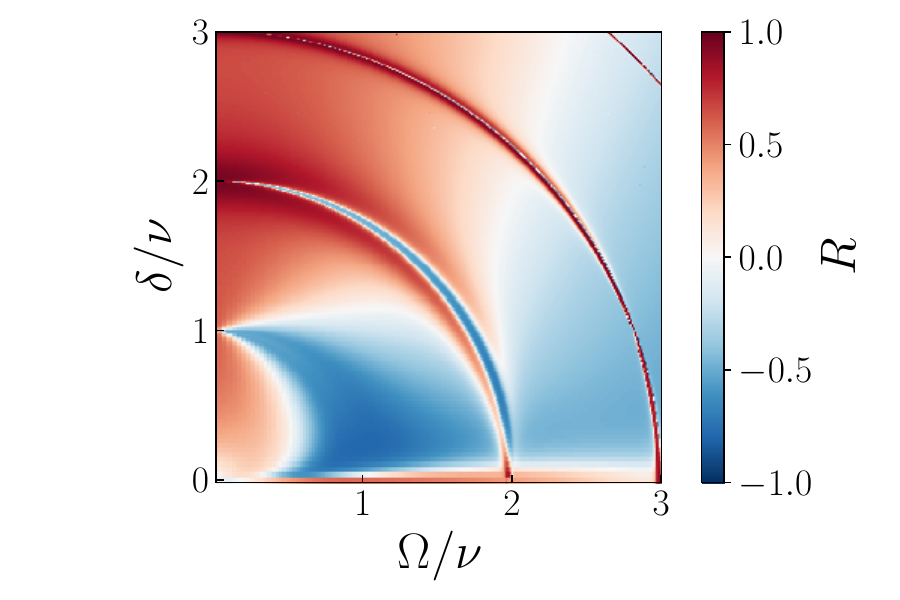}
    \caption{(Color online) Rectification factor $R$ as a function of the detuning $\delta$ and the coupling strength $\Omega$.
The hot and cold reservoirs were set, respectively, at temperatures $T_H = 5\nu$ and $T_C = 0.5 \nu$}
    \label{fig:rect2d}
\end{figure}

To gain a deeper understanding of how rectification depends on the parameters controlled by the laser properties, we once again fix the dissipation rates as $\gamma_M = \gamma_E = \gamma$ and present a simulation of the rectification factor for various values of $\delta$ and $\Omega$, as shown in Fig.~\ref{fig:rect2d}. Again, circular patterns tend to emerge across the parameter space, but now the phenomenology is more complex.
We observe a tendency toward negative rectification factors near the first circular sector. This trend appears in the second sector as well, though with sharper variations in $R$ close to the circular boundary.
When approaching the third circular sector, however,
the step size used in the numeric discretization
did not have enough resolution for us to make a precise observation.
In fact,
as it can also be observed in the other figures,
it appears that the variations in the features studied here
become increasingly sharper for the outer circular sectors.
This can ultimately constrain numerical studies
for large integers $m$ in Eq.~(\ref{circular}),
since finer and finer discretization steps will be needed.

\section{Final remarks} \label{sec:conclusion}

We investigated quantum heat transport through a trapped ion in various regimes of the coupling strength $\Omega$ and ion-laser detuning $\delta$. The heat current was driven by one reservoir coupled to the electronic part of the ion and another coupled to the motional part. Given that we considered couplings well into the strong regime, the use of the dressed master equation (DME) formalism was necessary. The current was calculated for the steady state of the DME.

We found that the heat current forms an intriguing circular pattern in the $(\delta, \Omega)$ space. The current maxima fall close to the circles $\delta^2 + \Omega^2 = (m \nu)^2$, for an integer $m$, where $\nu$ is the trap frequency. Furthermore, we observed that the coherence in the steady state, calculated with the decoupled basis, exhibits similar circular patterns, but with a notable distinction: it is not always the maximum that aligns with the circular lines. When the hot reservoir is coupled to the motional part, we observed that current maxima correlate with sudden drops in the residual coherence. Conversely, when the hot reservoir is coupled to the electronic part, we observed sudden peaks of leftover coherence when the current was maximal.

Other properties we investigated included differential heat conductivity and heat rectification, where we observed a rich phenomenology as the coupling strength $\Omega$ was varied. This may inspire applications such as thermal rectifiers and thermal diodes within this pivotal quantum technology setup. Finally, our study lays the foundation for further investigations into heat transport across controlled quantum systems.

\section*{Acknowledgements}
T. T. acknowledges financial  support from Coordena\c{c}\~ao de Aperfei\c{c}oamento de Pessoal de N\'ivel Superior (CAPES, Finance Code 001).
F. B. and F. L. S. acknowledge partial support from Brazilian National Institute of Science and Technology of Quantum Information (CNPq INCT-IQ 465469/2014-0).
F. L. S. acknowledges partial support from  Funda\c c\~ao de Amparo a Pesquisa do Estado de S\~ao Paulo (FAPESP) Process No. 2021/14135-1, CNPq (Grant No. 305723/2020-0) and (Grant No. 313068/2023-2), and CAPES/PrInt (Grant No. 88881.310346/2018-01).

\end{document}